\documentclass[twocolumn,tightenlines,prd,floatfix,nofootinbib]{revtex4}

 \usepackage[dvips,final]{graphicx}
  \usepackage{amssymb}
   \usepackage{amsmath}
    \usepackage{amsfonts}
     \usepackage{epsfig}
      \usepackage{bm}

\newcommand{\GeV}{\textrm{GeV}}

\begin{document}

\title{Diffractive deep inelastic scattering from multiple soft gluon exchange in QCD}

\author{Roman Pasechnik}
\email{roman.pasechnik@fysast.uu.se}
\author{Rikard Enberg}
\email{rikard.enberg@physics.uu.se}
\author{Gunnar Ingelman}
\email{gunnar.ingelman@physics.uu.se}

\affiliation{Department of Physics and Astronomy, Uppsala
University, Box 516, SE-751 20 Uppsala, Sweden}

\begin{abstract}
Diffractive hard scattering is interpreted as the effect of soft
gluon exchanges between the emerging energetic quarks and the
nucleon's color field, resulting in an overall color singlet
exchange. Summing multiple gluon exchanges to all orders leads to
exponentiation and an amplitude in analytic form. Numerical
evaluation reproduces the precise HERA data and gives new insights
on the density of gluons in the proton.
\end{abstract}

\maketitle

Diffractive deep inelastic scattering (DDIS) in lepton--proton
collisions involves hard scattering events where, in spite of the
large momentum transfer $Q^ 2$ from the electron, the proton emerges
essentially unscathed with only a very small transverse momentum,
keeping almost all of its original longitudinal beam momentum. The
leading proton is well separated in momentum space, or  rapidity
$y=1/2 \, \ln(E+p_z)/(E-p_z)$, from the central hadronic system
produced from the exchanged virtual photon's interaction with the
proton. Thus, this new class of events is characterized by a 
\textit{large rapidity gap} (LRG) void of final state particles.

Diffractive deep inelastic scattering (DIS) was discovered by the
ZEUS and H1 experiments at HERA \cite{ddisexp}, but the first
discovery of such a hard diffraction process was in $p\bar{p}$
collisions by the UA8 experiment \cite{UA8}. These processes had
actually been predicted \cite{IS85} by combining Regge phenomenology
for low-momentum transfer (soft) processes in strong interactions
via pomeron exchange, with large-momentum transfer (hard) processes
based on perturbative QCD. By parametrizing the parton content of an
exchanged pomeron (or alternatively diffractive parton density
functions) it is possible to describe the HERA data. However, the
extracted parton densities are not universal, since when used to
calculate diffractive hard scattering processes in $p\bar p$
collisions at the Tevatron one obtains cross sections an order of
magnitude larger than observed.

As an alternative dynamical interpretation the Soft Color
Interaction (SCI) model was developed in Ref.~\cite{Edin:1995gi},
based on the assumption that the hard perturbative part of the
interaction is the same as in ordinary DIS. The large momentum
transfer means that the hard subprocess occurs on a spacetime scale
much smaller than the bound state proton and is thus ``embedded'' in
the proton. The emerging hard-scattered partons, therefore,
propagate through the proton's color field and may interact with it.
Soft exchanges will dominate, due to the large coupling and the lack
of suppression from hard gluon propagators. Therefore, the momenta
of the hard partons are essentially undisturbed, which is consistent
with the fact that soft long distance interactions do not affect
hard short distance ones. However, the exchange of color may change
the color charges of the emerging partons such that the confining
string-like field between them will have a different topology,
resulting in a different distribution of the final state of hadrons
produced from the string hadronization~\cite{Andersson:1983ia}. In
particular, a region in rapidity without a string will result in an
absence of hadrons there, i.e.\ a rapidity gap. This SCI model is
very successful in describing data \cite{sciresult}, but lacks a
solid theoretical basis.

Here, we present a new QCD-based model, which leads to effective
color singlet exchange and thereby to diffractive scattering. The
model is inspired by the success of the SCI model, and may be seen
as an explicit realization of the earlier attempt \cite{BEHI05} to
understand this soft gluon exchange in terms of QCD rescattering.

\begin{figure}[!tb]
 \centerline{\includegraphics[width=0.4\textwidth]{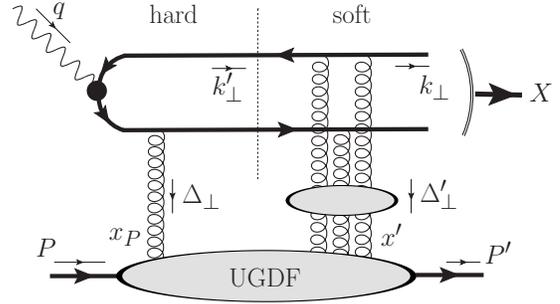}}
   \caption{\label{fig:amp}
   $\gamma^*p\to Xp$ process with resummed gluon exchanges, and illustration of the factorization in Eq.~(\ref{fac}) of the amplitude into a hard and a soft part connected via an unintegrated gluon density function (UGDF).}
\end{figure}

As depicted in Fig.~\ref{fig:amp}, the
exchanged photon fluctuates into a quark--antiquark color dipole
which interacts with the first gluon carrying a longitudinal
momentum fraction $x_P$. This defines the ``hard'' part of the
process. Both the $q\bar{q}$ dipole and the proton remnant then
have overall color octet charges and may interact through the
exchange of a number of soft gluons with longitudinal momentum
fractions $x_i'\ll x_P$, $\sum x_i'=x'$. This multiple gluon
exchange constitutes the ``soft'' part of the process and includes
at least one exchange. Thus the overall exchange from both the hard
and soft parts contains two gluons or more. The color factor of the
total soft exchange may combine to form an overall color singlet
together with the first, hard gluon. In this case both the
$q\bar{q}$-system and the proton remnant emerge as color singlets
and hadronize independently with a large separation in rapidity
due to the dominance of small-$x$ gluons in the proton. Lacking a
large momentum transfer to the proton, its remnant can recombine
into a leading proton with close to the full beam momentum. The
photon with a space-like virtuality $Q^2$ has through the momentum
exchange $x_P$ been turned into a time-like hadronic system $X$ of
invariant mass $M_X$.

In terms of the four-momenta $q$ of the photon, $P$ and $P'$ of the
initial and final proton, the important kinematical variables are
 \begin{equation}\label{vars}
 x_B=\frac{Q^2}{Q^2+W^2}\;,\quad \beta=\frac{Q^2}{Q^2+M_X^2}\;,\quad
 x_P=\frac{x_B}{\beta}\;,
 \end{equation}
where $Q^2=-q^2$ and
 \begin{equation}\label{vars1}
 M_X^2=\frac{1-\beta}{\beta}Q^2\,,\quad
 W^2\equiv (P+q)^2 = \frac{Q^2}{x_B}(1-x_B)\,.
 \end{equation}
In the forward limit of interest here, the momentum transfer $t=(P'-P)^2$ along
the proton line is small, $|t|\ll Q^2,\,M_X^2$.

Let us now outline the model and the calculation of the diffractive
structure function (see~\cite{nextpaper} for details).  
The involved momenta are specified in Fig.~\ref{fig:amp}.
We consider the asymmetric case where one hard gluon
carries most of the longitudinal momentum transfer $x_P$. Using
cutting rules, we derive a factorization of the amplitude into a
convolution of a hard part and a soft part. The hard part is treated
in normal perturbative QCD. The soft part consists of any number of
soft gluon exchanges, collectively in a color octet with total
$x'\ll x_P$. The soft exchanges are resummed, in the large $N_c$
limit, to all orders in $\alpha_s$. These soft gluons are not
perturbative, since the strong coupling becomes large. We model them
as interacting with quarks as perturbative gluons but with a
non-perturbative coupling to be specified below.

In the center-of-mass frame of the final state, i.e.\ the outgoing proton
with momentum $P'$, and the diffractive system $X$ with momentum
$q'=k_1+k_2$, where ${\bf k}_{1,\perp}=-{\bf k}_{2,\perp}\equiv {\bf
k}_{\perp}$, the quark virtuality
$k^2$ is the hard scale of the process,
$\mu_F^2\equiv k^2=\varepsilon^2+k_{\perp}^2$ and is expressed in terms
of its energy $\varepsilon$ and transverse momentum $k_{\perp}$ given by
 \begin{equation}
 \varepsilon^2=z(1-z)Q^2+m_q^2\,,\quad
 k_{\perp}^2=z(1-z)M_X^2-m_q^2\,, \label{k2}
 \end{equation}
where $z$ is the fraction of the longitudinal momentum carried by the
quark, and we consider the light quark mass limit $m_q\ll Q^2$.

In impact parameter space, with ${\bf b}$ conjugate to ${\bf \Delta}_{\perp}$,
the total amplitude for the $\gamma^*p\to Xp$ process can be written as a convolution 
of the hard and soft subprocess amplitudes and a function ${\cal V}$, which describes 
the distribution of gluons in the proton, 
 \begin{equation} \label{fac}
  M(\delta)\sim \int d^2b e^{-i{\bm
 \delta} {\bf b}} \hat{M}^\text{hard}\, \hat{M}^\text{soft}\, {\cal V}\, ,
 \end{equation}
where $\delta\equiv\sqrt{-t}=|{\bm \Delta}_{\perp}+{\bm \Delta}'_{\perp}|$. The function ${\cal V}$ is the Fourier transform of the unintegrated gluon distribution
function (UGDF), and will be specified below. 
This factorization is schematically illustrated in Fig.~\ref{fig:amp}.

The amplitude for the hard subprocess $\gamma^\star g\to q \bar{q}$
is decomposed into its longitudinal (L) and transverse (T) parts depending on the photon polarization.  
It includes the two possible couplings of the gluon to the $q{\bar q}$ pair, and can be
Fourier transformed to impact parameter space with ${\bf r}$ conjugate to
$k'_{\perp}$, the transverse momentum of a quark in the
intermediate state. The hard amplitudes are then given by
 \begin{align}
 \hat{M}_{L}^\text{hard} =& \, i{\cal C} \, \alpha_s(\mu_F^2) \sqrt{\beta} \, W^3
 z^{3/2}(1-z)^{3/2}\,
 K_0(\varepsilon r)\,, \label{MLhard}\\
 \hat{M}_{T,\pm}^\text{hard} =& \, i{\cal C} \alpha_s(\mu_F^2)
 \sqrt{\frac{2\beta}{1-\beta}}\,\frac{1}{\sqrt{x_P}} W^2
 z^{1/2}(1-z)^{3/2}\,
 \nonumber\\
\times&  \varepsilon K_1(\varepsilon r)  \frac{r_x\pm ir_y}{r}\,, \label{MThard}
 \end{align}
where ${\cal C}=8\pi e_q \sqrt{\pi \alpha_{em}}/N_c^2$ and $K_{0,1}$
are Bessel functions. 

We now turn to the soft subprocess amplitude, which can be
calculated  order-by-order and then resummed. It is important to
realize that these soft gluons carry nonperturbatively small
momentum transfers. We deal with this by taking $\alpha_s(\mu)$ at
very small scales as a parameter, which we fix using the
infrared-stable analytic perturbation theory
(APT)~\cite{Shirkov:1997wi}. In the limit that $\mu\to 0$, we use
$\alpha_s^\text{soft}\equiv\alpha_s^\text{APT}(\mu\to
\Lambda_\text{QCD})\simeq 0.7$. The softness of the color-screening
gluons with $x_i'\ll x_P$ implies that intermediate particles are
on-shell, and the dipole size $r$ is frozen. Cutting the
intermediate propagators, we pick up phase shifts originating from
the hard amplitude, which depend on the soft momentum exchanges
$\Delta'_{i,\perp}$. The diagram for one soft gluon exchange is a
tree-level diagram, while two-gluon exchange leads to a loop
integral. We calculate these contributions and perform the Fourier
transforms with respect to ${\bf \Delta}'_{\perp}$, where remarkably
the second order diagram turns out to be the second term in a series
that will exponentiate. This relies on the large $N_c$ limit, where
the color factors simplify to $C_F\simeq T_F N_c$. 
We get
 \begin{eqnarray}\nonumber
 &&e^{-i{\bf r}{\bf k}_{\perp}'}\hat{M}^\text{soft}_{1}=
 e^{-i{\bf r}{\bf k}_{\perp}}\,{\cal A}\,{\cal W}({\bf b},{\bf r})\,,\\
 &&e^{-i{\bf r}{\bf k}_{\perp}'}\hat{M}^\text{soft}_{2}=
 e^{-i{\bf r}{\bf k}_{\perp}}\,\frac{{\cal A}^2\,
 {\cal W}({\bf b},{\bf r})^2}{2!}\,,\quad \dots \label{resum}
 \end{eqnarray}
where  ${\cal A}=2\pi i\,C_F \alpha_s(\mu_{\text{soft}}^2)$,
$\mu_{\text{soft}}^2\sim\Delta_{\perp}^2$ is
the gluon virtuality, and we have defined
 \begin{eqnarray}
 {\cal W}({\bf b},{\bf r})=\frac{1}{2\pi}\ln\frac{|{\bf b}-{\bf r}|}
 {|{\bf b}|}\,.
 \end{eqnarray}
Summing over the number of soft gluons
in the final state leads to exponentiation in impact
parameter space, so that for the total soft subprocess amplitude we
finally get
 \begin{eqnarray} \label{M-soft}
 e^{-i{\bf r}{\bf k}_{\perp}'}\hat{M}^\text{soft}({\bf b},{\bf r})=
 -e^{-i{\bf r}{\bf k}_{\perp}}\,(1-e^{{\cal A}
 \,{\cal W}({\bf b},{\bf r})})\,.
 \end{eqnarray}
A similar expression was previously derived in the case of scalar
Abelian gauge theory in Ref.~\cite{Brodsky}. Note, that
$\hat{M}^\text{soft}({\bf b},{\bf r})$ is independent of the photon
polarization in the soft limit of small $\Delta'_{i,\perp}$.

To describe the coupling of the gluons to the proton, we use
the framework of $k_{\perp}$-factorization and generalized
(off-diagonal) UGDFs, which contain all information about the
non-perturbative coupling of the gluons to the proton, and is based
on a well-defined formal procedure for the transition from the
parton level to the hadron level (see e.g.\ Ref.~\cite{CF09}). The
coupling of a gluon to a quark is thus given by an off-diagonal
UGDF ${\cal
F}_g^{\text{off}}(x_P,x',\Delta_{\perp}^{2},{\Delta'_{\perp}}^{2},\mu_F^2)$,
absorbing a factor $C_F\alpha_s(\mu^2_{\text{soft}})/\pi$, and by
convention a gluon propagator $\Delta_\perp^{-2}$ into the UGDF in
order to keep it regular as $\Delta_\perp^2\to 0$. The absorbed
coupling $\alpha_s(\mu^2_{\text{soft}})$ corresponds to the coupling
of a screening gluon with virtuality
$\mu_{\text{soft}}^2\sim\Delta_{\perp}^2$ to a quark in the proton,
whereas the coupling of the hard gluon to the $q{\bar q}$ dipole and
to a quark in the proton is treated perturbatively at the
hard scale $\mu_F$.

Generalized parton distributions (GPDs) are not very constrained by data.
We use a prescription for the generalized UGDF, which was
introduced in Ref.~\cite{Pasechnik:2007hm}, motivated by positivity
constraints for GPDs \cite{Pire:1998nw}. This prescription works
well in the description of recent CDF data on central exclusive
charmonium production \cite{Pasechnik:2009qc}, and
allows incorporating the dependence on the longitudinal momentum fraction
and transverse momentum of the soft gluons in an explicitly symmetric way,
 \begin{equation}
 {\cal F}^{\text{off}}_g\simeq
 \sqrt{ {\cal F}_g(x_P,\Delta_{\perp}^2,\mu_F^2)
 {\cal F}_g(x',{\Delta_{\perp}'}^2,\mu_{\text{soft}}^2)}\,, \label{sqrt}
 \end{equation}
which explicitly involves the soft $x'$ dependence. Here ${\cal
F}_g$ is the normal diagonal UGDF, which depends on the gluon
virtuality, and which when integrated over this virtuality reduces
to the well-known collinear gluon PDF $g(x,\mu^2)$. The dependence
of ${\cal F}_g$ on the virtuality is not theoretically well-known
for small virtualities, and the UGDF is here modeled using the
collinear gluon PDF together with a simple Gaussian Ansatz for the
intrinsic transverse momentum dependence,
 \begin{align}
 \sqrt{x_P}{\cal F}^{\text{off}}_g &\simeq \sqrt{x_Pg(x_P,\mu_F^2)
 \, x'g(x',\mu_{\text{soft}}^2)}\,
 f_{G}(\Delta_{\perp}^2),\nonumber\\
 &f_{G}(\Delta_{\perp}^{2})={1}/({2\pi\rho_0^2})\,\exp\left({-{\Delta_{\perp}^2}/{2\rho_0^2}}\right),
 \label{ugdf}
 \end{align}
where the factor $\sqrt{x_P}$ is absorbed from the hard subprocess,
and the Gaussian width $\rho_0$ is the soft hadronic scale,
corresponding to the transverse proton size $r_p\sim 1/\rho_0$. Note
that this leads to an exponential $t$-dependence of the cross
section $\sim \exp(B_D t)$ with the diffractive slope $B_D =
1/\rho_0^2\simeq 6.9 \pm 0.2$~GeV$^2$~\cite{Chekanov:2008fh}.

The second PDF in Eq.~(\ref{ugdf}), associated with the soft gluon,
is evaluated at very low scale and very small $x'$. For this PDF we
can here introduce a function $\bar{R}_g(x',\mu_{\text{soft}}^2)$
which is assumed to be slowly dependent on $x'$ in the case $x'\ll
x_P$:
 \begin{eqnarray}
 \sqrt{x_P}{\cal F}^{\text{off}}_g \simeq \bar{R}_g(x',\mu_{\text{soft}}^2)
 \sqrt{x_Pg(x_P,\mu_F^2)}\,
 f_{G}(\Delta_{\perp}^2).
 \label{ugdfsat}
 \end{eqnarray}
The factor $\bar{R}_g$, therefore, contains all the soft physics
related with soft gluon couplings to the proton. It is interpreted
as the square root of the gluon PDF at very small $x'\ll x_P$ and
some soft scale $\mu_{\text{soft}}^2$. This is a non-perturbative
object, which contributes to the overall normalization and can be
determined from data. The factor $\bar{R}_g$ in Eq.~(\ref{ugdfsat})
is analogous to the skewedness parameter $R_g\simeq
1.2-1.3$, which accounts for the single $\log Q^2$ skewedness effect in
off-diagonal UGDFs \cite{Shuvaev:1999ce}. As we will see below, the
prescription (\ref{ugdfsat}) is consistent with the HERA data for
all available $M_X^2$ and $Q^2$.

The model (\ref{ugdf}) will lead to a \textit{linear} dependence of
the diffractive structure function on the gluon PDF, as compared to
the quadratic dependence often encountered in two-gluon exchange
calculations of DDIS~\cite{WM99}. This linear dependence is the same
as in the SCI model, where a linear dependence describes both
diffractive and non-diffractive events, and indicates a continuous
transition between the two types of events.

In terms of the UGDF in Eq.~(\ref{ugdfsat}), the factor
${\cal V}={\cal V}({\bf b},{\bf r})$ of Eq.~(\ref{fac}) is given by
\begin{align}
 {\cal V}({\bf b},{\bf r}) &=  \frac{1}{\alpha_s(\mu_{\text{soft}}^2)}
 \int\frac{d^2\Delta_{\perp}}{(2\pi)^2}\, \sqrt{x_P} \,
 {\cal F}_g^\text{off} \nonumber\\
 &\times \left\{e^{-i{\bf r}{\bf \Delta}_{\perp}}- e^{i{\bf r}{\bf
 \Delta}_{\perp}}\right\}
 e^{i{\bf b}{\bf \Delta}_{\perp}}. \label{V-UGDF}
 \end{align}

Straightforward calculations \cite{nextpaper} lead to the final
expressions for the diffractive structure functions
  \begin{align}
x_PF_L^{D(4)} &= {\cal S}\,Q^4M_X^2
\int_{z_{min}}^{\frac12}dz (1-2z)\, z^2(1-z)^2 |J_L|^2 \label{FL}\\
x_PF_T^{D(4)} &= 2{\cal S}\, Q^4\int_{z_{min}}^{\frac12}dz (1-2z) \left\{(1-z)^2+z^2\right\} |J_T|^2 \label{FT}
 \end{align}
where ${\cal S}={\sum_q e_q^2}/({2\pi^2N_c^3})$ sums over light
quark charges $e_q$, and
 \begin{align} \nonumber
 J_L=i\alpha_s(\mu_F^2)
 \int d^2{\bf r}d^2{\bf b}\, e^{-i{\bm \delta} {\bf b}}
 e^{-i{\bf r}{\bf k}_{\perp}}\,K_0(\varepsilon r) \nonumber \\
 \times\,{\cal V}({\bf b},{\bf r})
 \Big[1-e^{{\cal A}{\cal W}}\Big], \nonumber \\
 J_T=i\alpha_s(\mu_F^2)
 \int d^2{\bf r}d^2{\bf b}\, e^{-i{\bm \delta}
 {\bf b}}e^{-i{\bf r}{\bf k}_{\perp}}\,\varepsilon K_1(\varepsilon r)\nonumber \\
 \times\,\frac{r_x\pm ir_y}{r}  {\cal V}({\bf b},{\bf r})
 \Big[1-e^{{\cal A}{\cal W}}\Big]\,. \label{JLT}
 \end{align}

Let us briefly discuss the role of higher-order QCD corrections
in the framework of our model. One should distinguish corrections
from the hard gluon emission due to radiation from the partons in
the hard part, and corrections due to interactions between soft
gluons in the soft part.

In the first case, additional $s$-channel gluons emitted from the
hard scattering part can be described by DGLAP evolution. They play
an important role for large invariant masses $M_X^2\gg Q^2$, and
will be considered below.

In the second case, all interactions between soft gluons in the soft
scattering part are absorbed into the UGDF and thus into the soft
$\bar{R}_g$ factor, which enters the overall normalization. Only the
number of soft gluon legs attaching to the proton in the lower part
of the diagram, and how they attach to the partons in the upper part
are important. All long-distance interactions between the gluons are
treated as part of the color background field in the proton, and do
not affect the resummation procedure.

One could also imagine interactions between the gluon from the hard
part and one of the soft gluons. Such interactions contribute only
in the symmetric case $x'\sim x_P$, when all exchanged gluons are
either soft or hard. The first case is unrealistic as it may
happen only in the case of very small $M_X$ and $Q$ where QCD
factorization does not apply. The second case is suppressed by a 
small $\alpha_s$.
Such exchanges may be enhanced by large logarithms, leading to
exchange of a BFKL pomeron at non-zero momentum transfer $t$. This
is, however, a process with different kinematics and does not
contribute to forward diffraction.

In the large-$M_X$, or $\beta\to 0$, limit the additional emission
of a gluon in the final state becomes important. This is dominated
by the emission of a collinear gluon from the hard gluon, an
emission which is enhanced by a large logarithm. Such a gluon will
be well-separated from the $q\bar q$ pair in momentum space, and
will therefore contribute to building up a large $M_X$. We take this
contribution into account through a gluon splitting, described using
the DGLAP splitting function $P_{gg}$ as (see
e.g.~\cite{Ellis:1991qj})
\begin{eqnarray} \label{FDqqg}
x_PF_{q{\bar q}g}^{D(4)} \simeq \frac{1}{N_c^2}
\int\frac{dt_gdz_g}{t_g+m_g^2}\,{P}_{gg}(z_g)\frac{\alpha_s(t_g)}{2\pi}
x_PF_{q{\bar q}}^{D(4)}\,,
\end{eqnarray}
where $t_g$ is the gluon propagator and the integral is cut-off in
the infrared by the effective gluon mass
$m_g\simeq\Lambda_\text{QCD}$. The factor $N_c^{-2}$ appears because
the emitted gluon must contribute to the color singlet $X$ system.

The HERA data on the diffractive structure function
\cite{Chekanov:2008fh} are given in terms of the reduced cross
section,
 \begin{eqnarray*}
 x_P\sigma_r^{D(3)} = x_PF_{q{\bar q},T}^{D(3)}+
 \frac{2-2y}{2-2y+y^2}\,x_PF_{q{\bar q},L}^{D(3)}+x_PF_{q{\bar q}g}^{D(3)}
 \end{eqnarray*}
where $F^{D,(3)}_{L,T}(x_P,Q^2,\beta)$ is the diffractive structure function
integrated over $t$, the kinematical variable $y=Q^2/sx_B\leq 1$, and the
center-of-mass energy of $ep$-collisions at HERA is
$\sqrt{s}=318$~GeV.

\begin{figure*}[tb]
 \centerline{\includegraphics[width=0.9\textwidth]{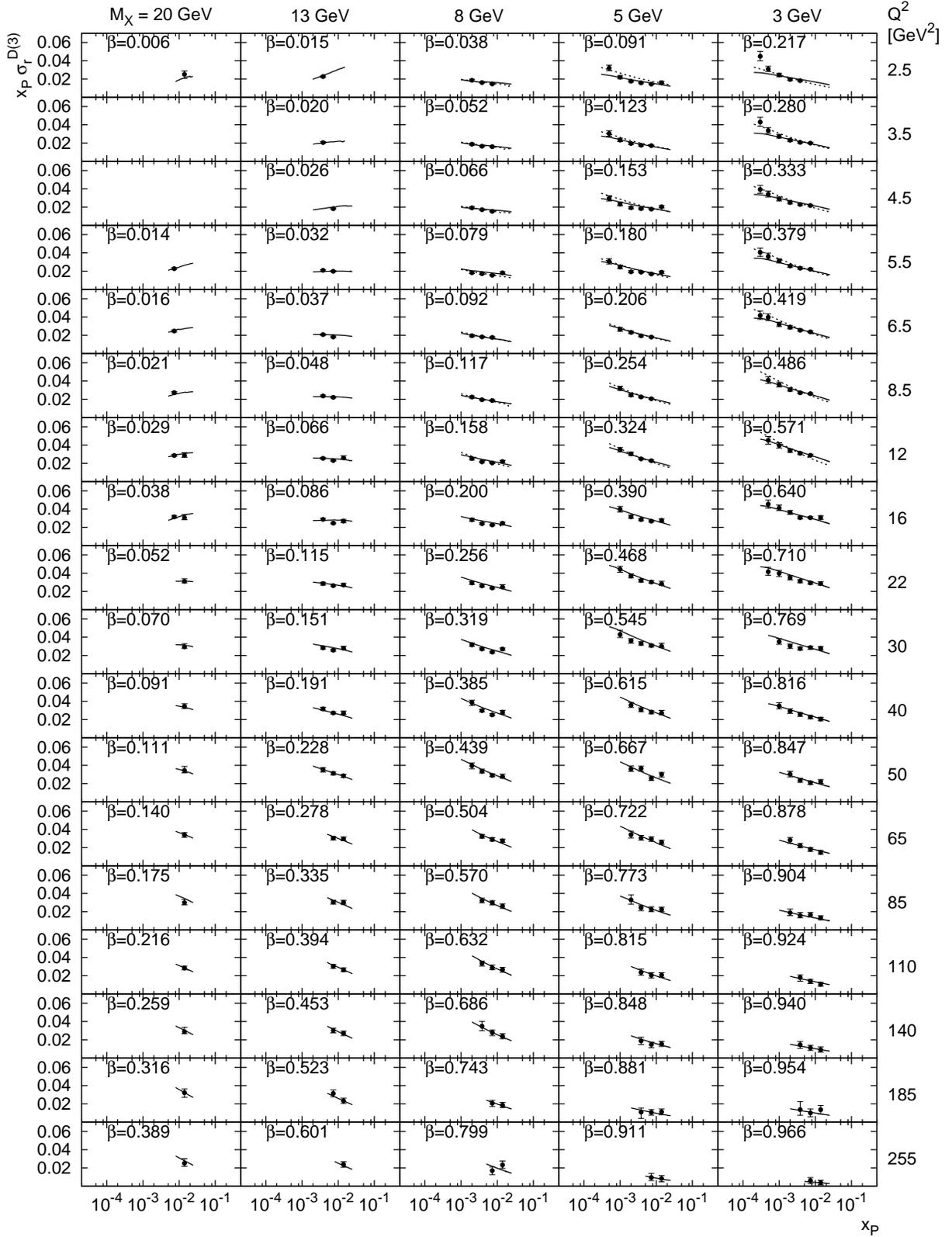}}
   \caption{\label{fig:data}
   The reduced cross section $x_P\sigma_r^{D(3)}(x_P,\beta,Q^2)$
   as a function of $x_P$ for different values of $M_X$ and $Q^2$.
   The latest ZEUS data \cite{Chekanov:2008fh}, from diffractive deep
   inelastic scattering events with a large rapidity gap, compared
   with our model using the CTEQ6L1 (full line) and GRV94 (dotted line)
   parametrizations of the gluon density in the proton.
   }
\end{figure*}

In Fig.~\ref{fig:data} we show the comparison of the results of our
model with the latest HERA data \cite{Chekanov:2008fh} on the
reduced cross section $x_P\sigma_r^{D(3)}(x_P,\beta,Q^2)$ as a
function of $x_P$ in bins of $\beta$ and $Q^2$. The figure shows our
main result, calculated using the CTEQ6L1 gluon
PDF~\cite{Pumplin:2002vw}, and also curves obtained using the older
GRV94 PDF~\cite{Gluck:1994uf}. The minimal factorization scale
$\mu_F$ is fixed to be $\mu^2_{F,min}=0.2\,\GeV^2$, which, together
with Eq.~(\ref{k2}) implies a minimal possible fraction of the quark
longitudinal momentum $z_\text{min}$ in the integrals in
Eqs.~(\ref{FL},\ref{FT}).

In these results, transverse polarization dominates in all bins. We
find that the $q{\bar q}$ contribution alone is enough to describe
all the data for $\beta\gtrsim 0.2$, below which the $q{\bar q}g$
contribution becomes significant.

The fixed parameters in our model, which all take reasonable
physical values, are an effective gluon mass, $m_g\simeq
\Lambda_\text{QCD}$, used to regulate the infrared divergence in the
$q\bar q g$ contribution, Eq.~(\ref{FDqqg}), and the soft coupling
constant $\alpha_s(\mu_\text{soft}^2)\simeq 0.7$.

 We also fit two physical quantities:\ the soft factor
$\bar{R}_g$, which absorbs the non-perturbative couplings of the
soft gluons to the proton, and the constituent quark mass $m_q$. For
our model to be consistent, these two parameters should not depend
strongly on the two large scales in the process, $M_X$ and $Q$.

We have found that 
$\bar{R}_g$ is close to unity for a wide range of scales. It
does not depend at all on $M_X$, and only in the lowest $Q$ bins
is there a noticeable increase in the normalization $\bar{R}_g$ by
at most a factor 4. Here, however, QCD factorization becomes
questionable and the conventional gluon PDFs are poorly known such
that using the GRV94 PDF instead of CTEQ6L1 this increase is
essentially removed~\cite{nextpaper}. Thus $\bar{R}_g$
is essentially an overall normalization factor close to unity, which 
contains unknown information on the density of soft gluons in the proton.

The kinematics, e.g. Eq.~(\ref{k2}), depends on an effective quark
mass corresponding to a dynamic, dressed quark generated dominantly
by softer gluon radiation which cannot be calculated theoretically.
Therefore, the parameter $m_q$ is fitted to data and found to have
only a slow variation with the large scales $M_X$ and $Q$; namely in
the range 0.3--1.2 GeV consistent with mainly soft dynamics, as
expected.

The uncertainty in the PDFs is illustrated in Fig.~\ref{fig:data},
where we show, in the bins of small $Q^2$ and $M_X$, the results
obtained using the GRV94 gluon PDF. In these bins GRV does better in
reproducing the data, because of its larger gluon density at these
small $x$ and scales.

In summary, in this Letter, we have presented a new QCD-based model
of soft gluon exchanges in the final state, which describes data on
the diffractive structure function very well. The model is inspired
by the phenomenologically successful Soft Color Interaction
model~\cite{Edin:1995gi} and on the work on such soft rescattering
in DIS in Ref.~\cite{Brodsky,BEHI05}. The full details of the model
and the calculations are presented in~\cite{nextpaper}.

We have considered diffractive DIS, as this is where the most
precise data are available, but the soft gluon exchanges arise due
to the proton's color field and should thus be of a universal
nature. Our model should therefore be applied to other processes,
for example, diffraction in hadron--hadron collisions or diffractive
vector meson production. It should also have effects on other
observables, such as the underlying event at LHC, which is
 important to understand and describe. This is also borne
out by the results~\cite{sciresult} from the SCI model, which has
not only been able to describe all diffractive data from HERA and
the Tevatron, but also other, non-diffractive data.

This work was supported by the Swedish Research Council and the Carl
Trygger Foundation. We are grateful to Igor Anikin for valuable
discussions.

\end{document}